# A Comparison of O(1) and Cyrus-Beck Line Clipping Algorithms in $E^2$ and $E^3$

Václav Skala, Pavel Lederbuch, Bohumír Sup[1]
Department of Information Technology and Computer Science
University of West Bohemia
Univerzitní 22, Box 314
306 14 Plzeò
Czech Republic

skala@kiv.zcu.cz         http://yoyo.zcu.cz/~skala
lederbuc@kiv.zcu.cz      http://ody.zcu.cz/~lederbuc
sup@kiv.zcu.cz           http://yoyo.zcu.cz/kiv_info/sup_e.html

**Abstract**

A comparison of a new algorithm for line clipping in $E^2$ and $E^3$ by convex polygon and/or polyhedron with O(1) processing complexity and Cyrus-Beck algorithm is presented. The new algorithm in $E^2$ is based on dual space representation and space subdivision technique. The principle of algorithm in $E^3$ is based on the projection of polyhedron to three orthogonal $E^2$ coordinate systems. Algorithms have optimal complexities O(1) and demonstrates that preprocessing can be used to speed up the line clipping significantly. Obvious applications are for one polygon and/or polyhedron and many clipped lines. Detailed theoretical estimations and experimental results are also presented.

**Symbols:**

| | |
|---|---|
| $E^2$ | an Euclidean space |
| $D(E^2)$ | dual representation of an Euclidean space |
| x, y | point coordinates in $E^2$ |
| a, b, c | line coefficients in $E^2$ |
| k, q, m, p | line coefficients in $E^2$, point coordinates in semidual space |
| P | polygon and/or polyhedron |
| N | number of edges and/or facets |
| M | number of clipped lines |
| Pr | probability of intersection of a polygon by clipped line |
| r | line to be clipped |
| $n_k, n_q, n_m, n_p$ | number of steps for subdivision in the specified directions |
| $T_{CB}$ | processing time of CB algorithm |
| $T_{O(1)}$ | processing time of a new algorithm |
| $T_{prep}$ | preprocessing time of a new algorithm |
| $\nu_1, \nu_2$ | efficiency coefficients |

[1] Supported by the grant UWB-156

# 1. Introduction

Many algorithms for line clipping in $E^2$ and $E^3$ were developed, see [Ska94a], [Ska95a]. Algorithms for line clipping are mostly based on the Cyrus-Beck (CB) algorithm and its modifications. The aim of the line clipping algorithm is to find a part of the given line which is inside of the given polygon and/or polyhedron. Algorithms for line clipping are mostly restricted to line clipping against convex polygon and/or polyhedron. Since the line clipping problem solution is a bottleneck of many packages and applications it is convenient to use the fastest algorithm. Algorithm comparisons and description can be found in [Kol94], [Ska94a], [Ska94b], [Ska95a]. Those comparisons included algorithms with algorithm complexity between O(N) and O(1). In the tested algorithm we are using the pre-processing for speeding up problem solution, it decreases algorithm processing complexity. We are using algorithm for two dimensional line clipping with expected O(1) complexity as next simplification. Acceleration of algorithm processing is dependent on the memory consuming.

# 2. Dual space representation

Any line $r \in E^2$ can be described by an equation
$$ax + by + c = 0$$
and it can be rewritten as
$$y = kx + q, \quad |k| \neq \infty$$
or
$$x = my + p, \quad |m| \neq \infty$$

It means that the line $r \in E^2$ can be represented using asymmetrical model of space representation as a point $D(r) = [k,q] \in D(E^2)$ or $D(r) = [m,p] \in D(E^2)$, see Fig. 2.2. This representation model has very interesting features and usage that can be found in [Sto89], [Nie95], [Kol94], [Zac95], [Zac96]. Generally it is possible to show the relation between fundamental geometric primitives by the Table 2.1. In the following we will consider situations in $E^2$ only.

| Space | Euclidean rep. | Dual representation |
|---|---|---|
| $E^2$ | line | point |
|  | point | line |
| $E^3$ | plane | point |
|  | line | line |
|  | point | plane |

**Table 2.1.** Representation of geometrical primitives

It can be shown that a polygon $P \in E^2$, see Fig. 2.1.a, can be represented by an infinite area in dual space D($E^2$).

The **test** whether a line $r \in E^2$ intersects a convex polygon $P \in E^2$ is dual to the well known Point-in-Polygon test. Algorithms for Point-in-Polygon test usually have *O(N)* or *O(log₂ N)* complexities. Solution of the line clipping problem in $E^2$ generally consists of two steps:
- test whether a line intersects the polygon.
- selection of polygon edges which are intersected by the given line and computation of intersection points.

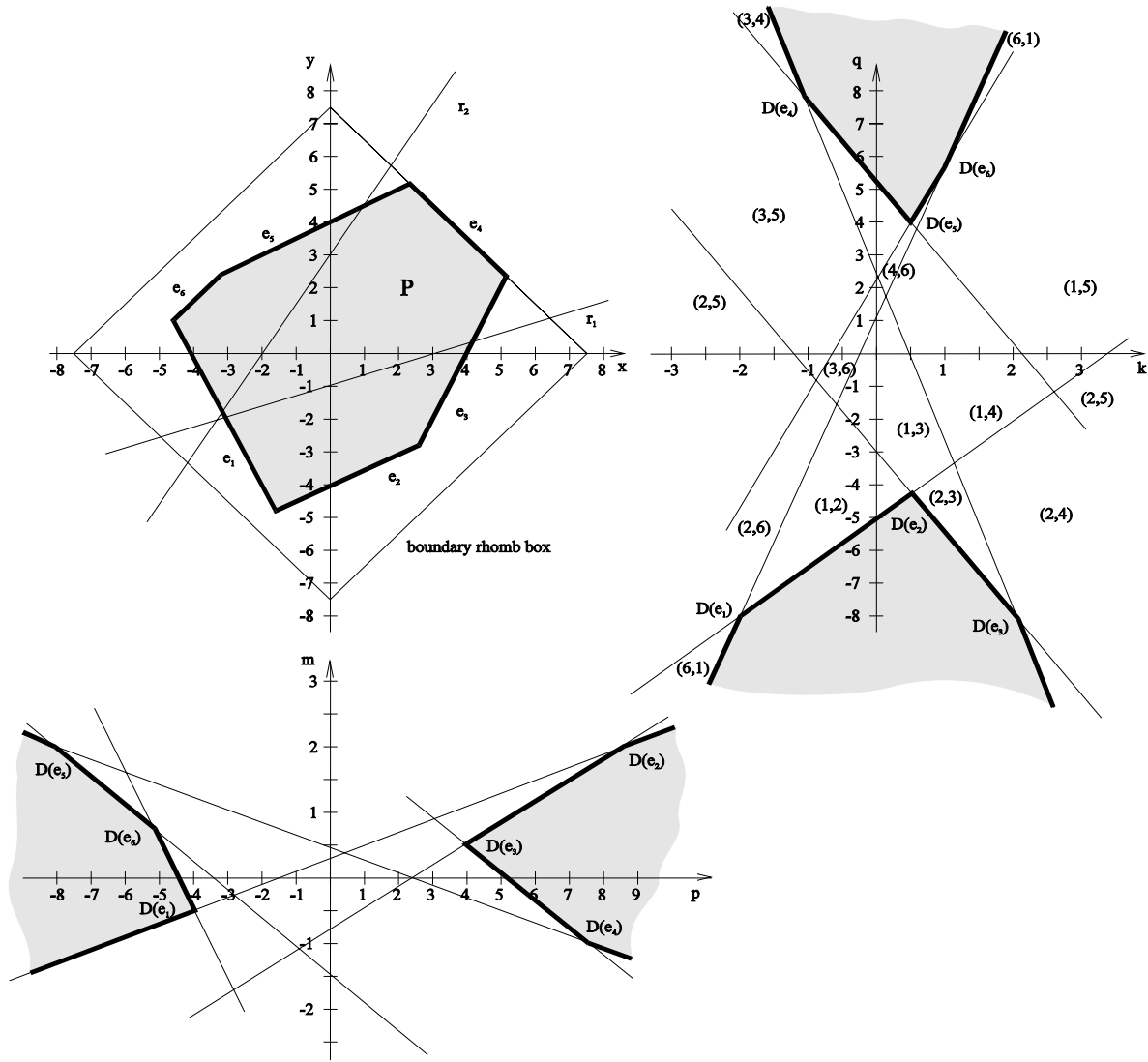

**Fig. 2.1.** Dual space representation of polygon in $E^2$

It means that the line clipping problem solution is more complex than the Point-in-Polygon test. Nevertheless, the $O(log_2 N)$ complexity of the Point-in-Polygon test leads to the new $O(log_2 N)$ line clipping algorithm development, see [Ska94a].

But there are two problems that must be solved, when dual space representation is used:
- zones in dual space are **infinite** and it is difficult to represent them,
- it is necessary to find fast method for determination in which zone the point $D(r)$ lies.

Let us consider a modified boundary rhomb box so that the given polygon is inside of a rectangle, Fig. 2.2. It can be seen that $q$ and $p$ values of lines which intersects the polygon are limited.

The given line $r \in E^2$ can be represented as
$$y = kx + q \quad \text{if} \quad |k| \leq 1$$
resp.
$$x = my + p \quad \text{if} \quad |m| < 1$$

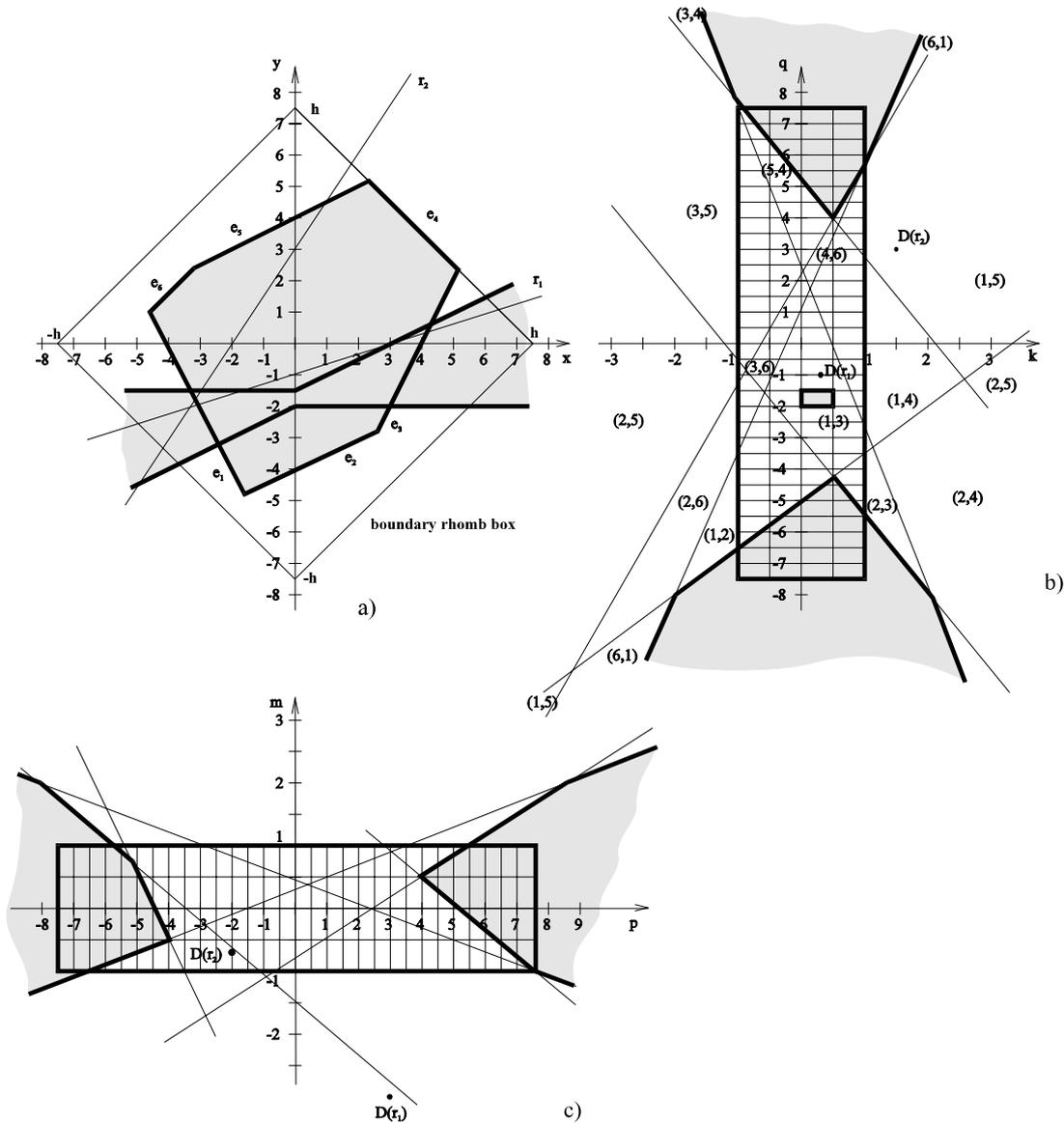

**Fig. 2.2.** Semidual space representation

If this representation is used then *k* and *m* values are **limited**. Then values *[k, q]*, resp. *[m, p]* are from the limited area $<-1,1>$ for *k* and *m* and $<-h,h>$ [2] for *q* and *p*. We will denote those two limited spaces as **semidual spaces**, see Fig. 2.2.

Several sophisticated techniques for detection in which zone a point *D(r)* lies have been developed as a part of computational geometry, see [Pre85a]. One possibility is to use the space subdivision technique. If semidual spaces for *(k,q* or *(m,p)* are subdivided into small rectangles, it is possible to pre-compute Active Edge List (AEL) of polygon edges that interfere-in semidual space with the given rectangle. If rectangles are small enough then each AEL will contain only two polygon edges. Each rectangle in the semidual space representation corresponds to an infinite „butterfly" zone in $E^2$, see Fig. 2.2.

It is necessary to emphasise that the rhomb that bounds the polygon must be as small as possible. Generally the limits for *q* and *p* axis can be different. It decreases the memory requirements significantly.

---

[2] ***h*** are values of y axis intersecting by rhomb box

## 3. Space subdivision in $E^2$

The space subdivision technique is used to detect the zone in which a point $D(r)$ lies. Semidual spaces for *(k, q)*, resp. *(m, p)* are subdivided into small rectangles. Each rectangle is a dual representation of a zone in $E^2$, see Fig. 2.2a. For each zone in $E^2$ is possible to compute list of polygon edges that interfere with it. The list of those edges is named Active Edge List (AEL), see [Ska94c] for details. It is necessary to point out that number of members of AEL depends on the geometric shape of the given polygon and also on the number of subdivision steps in *(k, q)*, resp. *(m, p)* spaces. If rectangles are small enough (it means that subdivision is high) then each list contains only two edges of the given polygon.

It is necessary to find a criterion for semidual spaces subdivisions so that just one pair of polygon edges is in the AEL.

For *(k, q)* semidual space we use the equation
$$y = kx + q$$
Then

$n_q > 2a / \Delta y$ where $\Delta y = \min\{|y_i - y_j|\}$ $i,j \in \langle 0,n \rangle$ & $i \neq j$ & $\Delta y > 0$

$n_k > 2 / \Delta k$ where $\Delta k = \min\{|k_i - k_j|\}$ $i,j \in \langle 0,n \rangle$ & $i \neq j$ & $\Delta k > 0$

Similarly for *(m, p)* semidual space
$$x = my + p$$
and

$n_p > 2a / \Delta x$ where $\Delta x = \min\{|x_i - x_j|\}$ for all $i, j$ & $i \neq j$ & $\Delta x > 0$

$n_m > 2 / \Delta m$ where $\Delta m = \min\{|m_i - m_j|\}$ for all $i, j$ & $i \neq j$ & $\Delta m > 0$

It means that the $n_k$ and $n_q$, resp. $n_m$ and $n_p$ values depend on geometric shape of the given polygon.

These conditions guarantee that each list of AEL contains up to three edges. It is necessary to point out that these conditions can extremely increase subdivisions of semidual spaces and memory requirements can be above system possibilities. That is why these conditions cannot be realized. Then it is essential to find an optimal level of subdivision which subdivide the semidual spaces sufficiently, but on the other hand which do not exceed above available memory. Experimental results of space subdivision for a polygon with N = 10 are presented in Fig. 3.1.,Fig. 3.2.

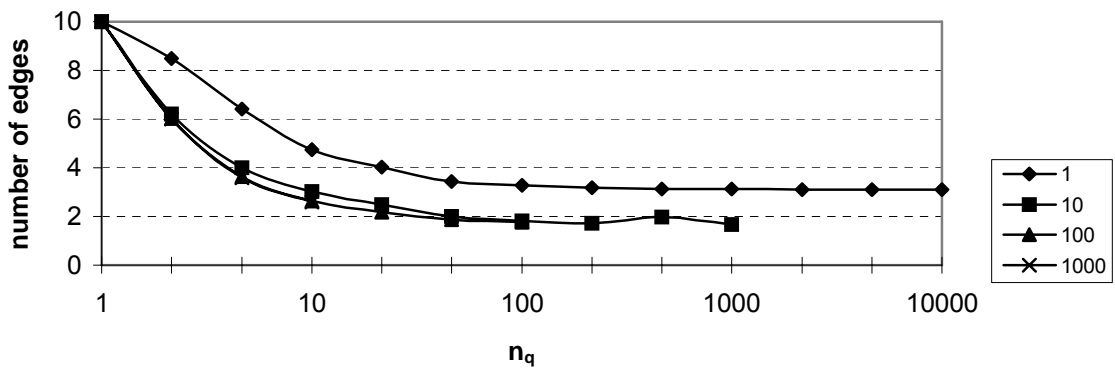

**Fig. 3.1.** Number of edges in AEL dependent on subdivision in the direction of q
(N = 10, $n_k$ is 1, 10, 100 and 1000)

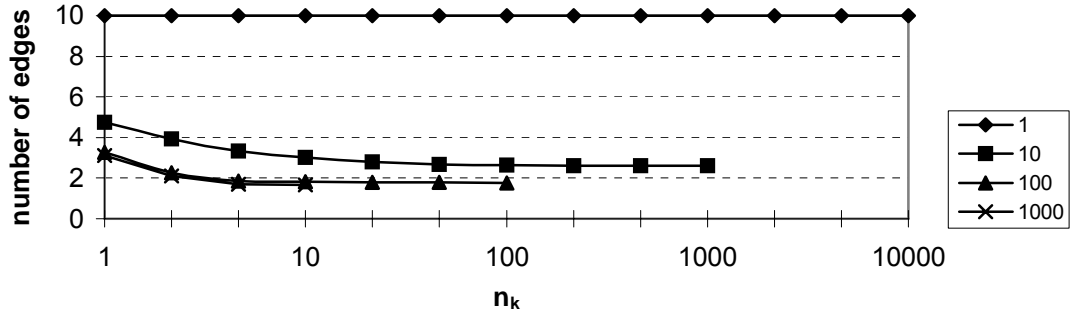

**Fig. 3.2.** Number of edges in AEL dependent on subdivision in the direction of k
(N = 10, $n_q$ is 1, 10, 100 and 1000)

The experimental results show that subdivision in the direction of q, resp. p is more significant than subdivision in the direction of k, resp. m. Figures 3.1. and 3.2. shows that adequate number of subdivision steps in q direction is 10 multiple N and in k direction is N.

The construction time of AEL is presented in Table 3.1. and Fig. 3.3.

| $n_q$ \ N | 3 | 5 | 10 | 20 | 50 |
|---|---|---|---|---|---|
| 1 | 0,00 | 0,00 | 0,06 | 0,10 | 0,22 |
| 2 | 0,06 | 0,05 | 0,05 | 0,11 | 0,28 |
| 5 | 0,05 | 0,06 | 0,11 | 0,28 | 0,55 |
| 10 | 0,11 | 0,11 | 0,28 | 0,49 | 1,15 |
| 20 | 0,17 | 0,27 | 0,44 | 0,87 | 2,20 |
| 50 | 0,44 | 0,66 | 1,20 | 2,25 | 5,39 |
| 100 | 0,88 | 1,32 | 2,36 | 4,45 | 10,76 |
| 200 | 1,75 | 2,59 | 4,67 | 8,90 | 21,48 |
| 500 | 4,40 | 6,49 | 11,70 | 22,14 | 53,61 |
| 1000 | 8,90 | 13,24 | 23,95 | 45,48 | 110,07 |

**Table 3.1.** Preprocessing time of AEL
(subdivision in the direction of k : $n_k$ = 10)

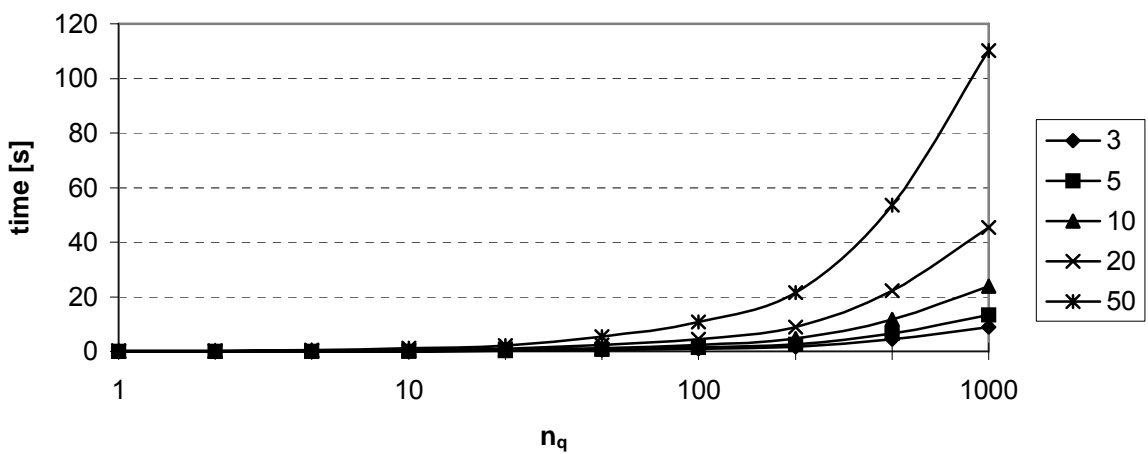

**Fig. 3.3.** Preprocessing time of AEL
(subdivision in the direction of k: $n_k$ = 10)

## 4. Theoretical consideration of O(1) algorithm in $E^2$

The O(1) algorithm has been tested and compared with the Cyrus-Beck algorithm. The reason for the choice of CB algorithm is that this algorithm is numerically very stable and its behaviour does not depend on geometric properties of the given polygon and clipped lines.

Let us consider that *N* is a number of edges of the given polygon. Theoretical complexity of CB algorithm, see [Ska93], is

$$T_{CB} = 590 + 621*N$$

Theoretical complexity of O(1) algorithm can be estimated as, see [Ska95b]

$$T_{O(1)} = 2020$$

Let us introduce algorithm efficiency coefficients as

$$v_1 = \frac{T_{CB}}{T_{O(1)}} \quad , \quad v_2 = \frac{T_{CB}}{T_{O(1)} + T_{prep}}$$

then expected efficiency of the O(1) algorithm is described by Table 4.1, see [Ska95b] for details.

| N | 3 | 4 | 5 | 10 | 50 |
|---|---|---|---|----|----|
| $v_1$ | 1.3 | 1.6 | 1.9 | 3.4 | 15.7 |

**Table 4.1.** Theoretical estimation of efficiency

## 5. Experimental results of O(1) algorithm and comparison with CB algorithm in $E^2$

Experimental results of processing time are presented in Table 5.1, Table 5.2 and Fig. 5.1.

| N | 3 | 4 | 5 | 10 | 20 | 50 |
|---|---|---|---|----|----|----|
| $T_{CB}$ | 0,99 | 1,26 | 1,59 | 3,13 | 6,59 | 15,38 |
| $T_{prep}$ | 0,44 | 0,55 | 0,66 | 1,20 | 2,25 | 5,39 |
| $T_{O(1)}$ | 0,50 | 0,50 | 0,50 | 0,54 | 0,55 | 0,55 |
| $T_{prep}+T_{O(1)}$ | 0,94 | 1,05 | 1,16 | 1,74 | 2,80 | 5,94 |
| $v_1$ | 2,0 | 2,5 | 3,2 | 5,8 | 12,0 | 28,0 |
| $v_2$ | 1,1 | 1,2 | 1,4 | 1,8 | 2,4 | 2,6 |

**Table 5.1.** Experimental results of processing times
(M=10.000, Pr=0%, coefficients of O(1): $n_k$=10, $n_q$=50)

| N | 3 | 4 | 5 | 10 | 20 | 50 |
|---|---|---|---|----|----|----|
| $T_{CB}$ | 0,99 | 1,32 | 1,65 | 3,13 | 6,15 | 15,32 |
| $T_{prep}$ | 0,44 | 0,55 | 0,66 | 1,20 | 2,25 | 5,39 |
| $T_{O(1)}$ | 1,32 | 1,32 | 1,32 | 1,32 | 1,32 | 1,49 |
| $T_{prep}+T_{O(1)}$ | 1,76 | 1,87 | 1,98 | 2,52 | 3,57 | 6,88 |
| $v_1$ | 0,8 | 1,0 | 1,3 | 2,4 | 4,7 | 10,3 |
| $v_2$ | 0,6 | 0,7 | 0,8 | 1,2 | 1,7 | 2,2 |

**Table 5.2.** Experimental results of processing times
(M=10.000, Pr=100%, coefficients of O(1): $n_k$=10, $n_q$=50)

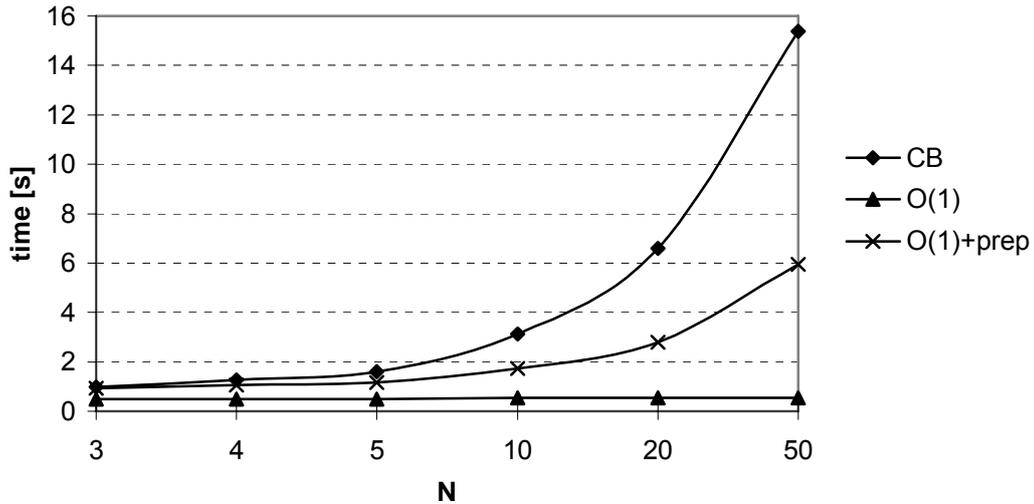

**Fig. 5.1.** Experimental results of processing times
(related to values in Table 5.1)

A comparison of theoretical estimations and experimental results is presented in Table 5.3.

| N | 3 | 4 | 5 | 10 | 50 |
|---|---|---|---|---|---|
| $\nu_1$ (theoret.) | 1,3 | 1,6 | 1,9 | 3,4 | 15,7 |
| $\nu_1$ (exp. Pr=0%) | 2,0 | 2,5 | 3,2 | 5,8 | 28,0 |
| $\nu_1$ (exp. Pr=100%) | 0,8 | 1,0 | 1,3 | 2,4 | 10,3 |

**Table 5.3.** Theoretical and experimental efficiencies

Dependence of a processing time of O(1) algorithm on a probability that clipped lines intersect the given polygon for N=10 is shown in Table 5.4 and Fig. 5.2. Experimental results show that O(1) algorithm is faster than CB algorithm if number of clipped lines is greater than 5000. The limit value is the same for all probabilities of intersection of a polygon by clipped line.

| M | 1000 | 2000 | 3000 | 4000 | 5000 | 10000 | 20000 | 50000 | 100000 |
|---|---|---|---|---|---|---|---|---|---|
| $T_{CB}$ | 0,28 | 0,60 | 0,88 | 1,21 | 1,48 | 3,02 | 5,99 | 15,05 | 30,10 |
| $T_{prep}$ | 1,20 | 1,20 | 1,20 | 1,20 | 1,20 | 1,20 | 1,20 | 1,20 | 1,20 |
| $T_{O(1)}$ | 0,06 | 0,11 | 0,16 | 0,22 | 0,27 | 0,50 | 1,04 | 2,58 | 5,16 |
| $T_{O(1)}+T_{prep}$ | 1,26 | 1,31 | 1,36 | 1,42 | 1,47 | 1,70 | 2,24 | 3,78 | 6,36 |
| $\nu_1$ | 4,7 | 5,5 | 5,5 | 5,5 | 5,5 | 6,0 | 5,8 | 5,8 | 5,8 |
| $\nu_2$ | 0,2 | 0,5 | 0,6 | 0,9 | 1,0 | 1,8 | 2,7 | 4,0 | 4,7 |

**Table 5.4.** Processing times for different number of clipped lines
Polygon with N = 10, coefficients of O(1) : $n_k$=10, $n_q$=50

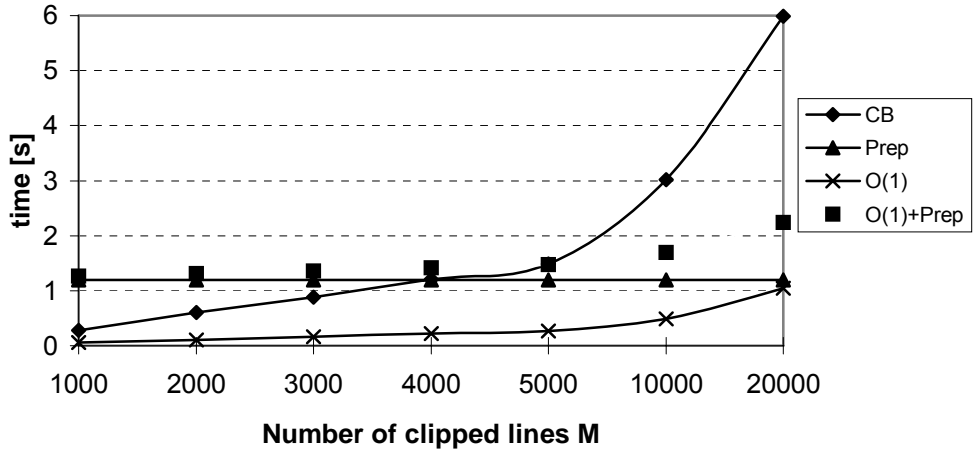

**Fig. 5.2.** Processing times for different number of clipped lines

The processing time of the proposed O(1) algorithm depends on a probability of intersection of the given polygon. It is also presented in Table 5.5 and Fig. 5.3. The time complexity of the CB algorithm is nearly constant, but the more clipped lines intersect the polygon, the more processing time is needed in O(1) algorithm. It is due to the test whether a line intersects a polygon is faster for a line does not intersecting the given polygon than for a line which intersects it.

| Pr | 0 | 10 | 20 | 30 | 40 | 50 | 60 | 70 | 80 | 90 | 100 |
|---|---|---|---|---|---|---|---|---|---|---|---|
| $T_{CB}$ | 3,02 | 3,08 | 3,02 | 3,07 | 3,08 | 3,07 | 3,13 | 3,07 | 3,08 | 3,14 | 3,19 |
| $T_{prep}$ | 1,20 | 1,20 | 1,20 | 1,20 | 1,20 | 1,20 | 1,20 | 1,20 | 1,20 | 1,20 | 1,20 |
| $T_{O(1)}$ | 0,54 | 0,60 | 0,66 | 0,71 | 0,82 | 0,88 | 0,99 | 1,05 | 1,21 | 1,21 | 1,32 |
| $T_{O(1)}+T_{prep}$ | 1,74 | 1,80 | 1,86 | 1,91 | 2,02 | 2,08 | 2,19 | 2,25 | 2,41 | 2,41 | 2,52 |
| $v_1$ | 1,7 | 1,7 | 1,6 | 1,6 | 1,5 | 1,5 | 1,4 | 1,4 | 1,3 | 1,3 | 1,3 |
| $v_2$ | 5,6 | 5,1 | 4,6 | 4,3 | 3,8 | 3,5 | 3,2 | 2,9 | 2,5 | 2,6 | 2,4 |

**Table 5.5.** Processing times for different probability of intersection
N = 10, M = 10.000, coefficients of O(1) : $n_k=10$, $n_q=50$

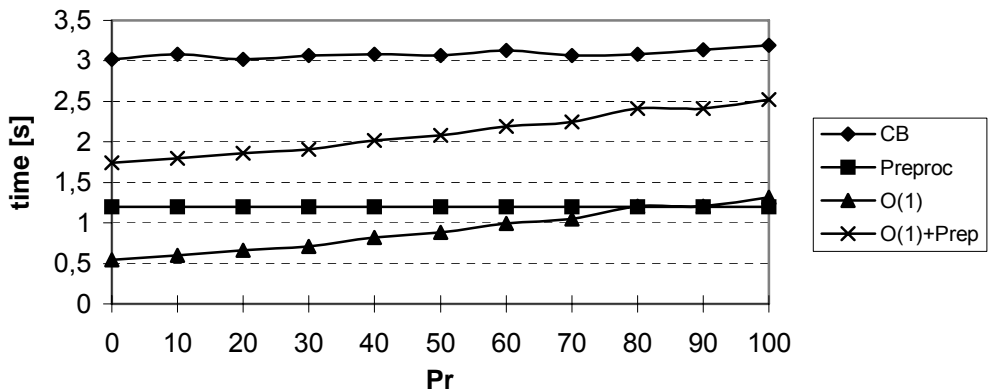

**Fig. 5.3.** Processing times for different probability of intersection

## 6. Principle of the proposed algorithm in E$^3$

In the Table 2.1 you can see, fundamental relation between geometric primitives. If we look at row „E$^3$", we cannot simplify our problem with using duality. In E$^3$ is dual representation of Euclidean line again the line. It is main reason for simplification. As we can see in the Table 2.1 in row „E$^2$" dual representation of Euclidean line is the point. With this simplification we can use Point-in-Polygon test algorithm in dual representations of polygon and lines. This algorithm is known with algorithm complexity O(1).

Let us assume that a convex polyhedron *P* is defined by triangular facets (generally it is not necessary). The triangular facets were used for simplification of problems with polygon construction and description.

Let us assume that the given polyhedron $P \in E^3$ is projected to the three orthogonal $E^2$ planes, see Fig.6.1 (only one projection is shown; only the front facets are shown). The planes are defined as *xy*, *xz* and *yz* where *x,y,z* are axes. If we use semidual space (see chapter 2) we have six E$^2$ representation of given polyhedron. This representations are:

|       | xy | xz | yz |
|-------|----|----|----|
| k,q   | 1  | 3  | 5  |
| m,p   | 2  | 4  | 6  |

**Table 6.1.**

The table show six planes with its numbers. On this number we will reference in algorithm 6.1.

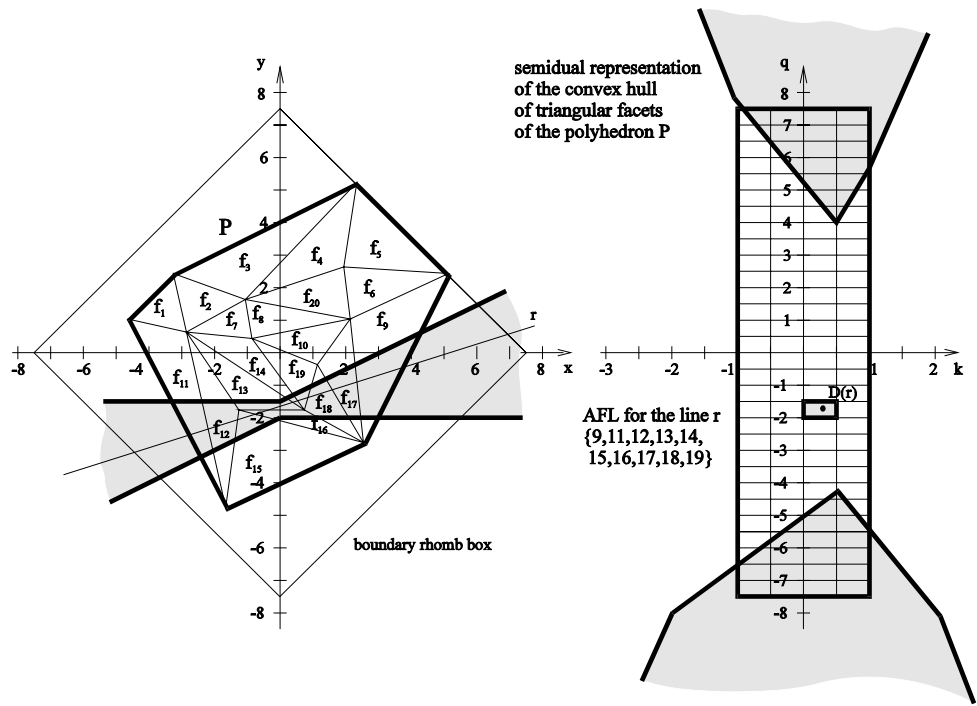

**Fig. 6.1.** Semidual representation of convex polyhedron

Let us assume the semidual representation for *[k,q]* values. Then the semidual space can be split into small rectangles using space subdivision technique. Each rectangle in semidual space represents an infinite „butterfly" zone in $E^2$ space. There are six Active

Facets Lists (AFL) of facets associated with each „butterfly" zone. The AFL contain information on all facets that interfere with the zone. The AFL can be represented as a list of pointers but such an implementation would be quite memory consuming as its length can be estimated as $(\sqrt{N})$.

It is necessary to point out that we must prepare both *(k,q)* and *(m,p)* semidual representations for all three planes $\rho_i^+$, $i = 1,2,3$. It means that we need six semidual representations altogether. For each clipped line *r* we must select two planes $\rho_{i_1}$ and $\rho_{i_2}$ and appropriate semidual representations, i.e. *(k,q)* or *(m,p)*, for each selected plane. The proposed algorithm is described by Alg.6.1.

We must select two planes $\rho_{i_1}$ and $\rho_{i_2}$, $i_1 \neq i_2$ for the given line $r \in E^3$. The criterion for selecting two planes we can derive from singular case. In some cases the line can be parallel or „almost parallel" with some axis. In these cases the line projection to the one of the three planes is wrong conditioned or does not exist. If our line has smallest angle with x axis, we select *xy* and *xz* projection planes.

**O(1) clipping algorithm:**
  **global constants**:   a - size of boundary rhomb box for the given polyhedron P,
                            $n_q$ - number of subdivision for q axis in semidual space representation,
                            $n_k$ - number of subdivision for k axis in semidual space representation,
                            $k_r, q_r$ - topical arguments value of given line *r* -
                                    for all spaces assume $n_q = n_p$, $n_k = n_m$, according to context.

c := 1;
**for** i := $i_1, i_2$ **do**  (* plane index i $\in$ {1,2,3} *)
**begin**
       **if** $|k_r| \geq 1$ **then** j := 2*i - 1 **else** j := 2*i; (* j is index of the AFL - see Table 3.1 *)
       ii := 2*a / $n_q$*$q_r$; jj := 2 / $n_k$*$k_r$;     (* index zone determination *)
       $\Omega_c$ := AFL$_j$[ii,jj]; c:= c+1;
**end**;
$\Omega := \Omega_1 \cap \Omega_2$;
**for** i := 1 **to** N **do** (* N - number of <u>selected</u> polyhedron facets *)
       **if** $\Omega[i] = 1$ **then**       (* i-th bit of AFL *)
              Detail $E^3$ Test (facet$_i$, r); (* computation is done usually for 4 - 6 facets only *)
--

**Algorithm 6.1.**

The condition $\Omega[i] = 1$ is true for 4-6 facets only. It is obvious that the algorithm complexity does not depend on the number of polyhedron facets but on the length of the final set $\Omega$. Function *Detail $E^3$ Test* is based on the CB algorithm[3] that is performed only for facets that are included in the final set $\Omega$. If the rectangles are small enough then 4 - 6 facets can be expected in the final set $\Omega$ nearly for all cases.

Because all steps in Alg.6.1 have *O*(1) complexity the whole algorithm has *O*(1) complexity, too. It is necessary to point out that number of members in AFL depends on subdivision in *(k,q)*, resp. *(m,p)* spaces and also on geometric shape of the given polyhedron,

---
[3] This test is practise for original polyhedron in $E^3$

see [Ska94c]. For more efficient algorithm of the last loop evaluation from Alg.6.1. see [Ska93].

E.g.: For polygon with 2112 facets and subdivision $n_k=n_q=n_m=n_p=15$ is average number of interfering facets about **290** (5000 lines tested). This number was counted before „∩" operation. The number of interfering facets is about **5** for final set Ω (after „∩" operation).

Because of that it is more convenient to use binary maps [Ska93]. This technique is based on a binary vector in which the i-th bit is set to „1" if the i-th object is in the AFL. Using this technique the memory requirements are small and the intersection operation is implemented as the bit-wise operation **and**, that is very fast in comparison with detail $E^3$ test.

## 7. Construction of AFL

An algorithm for setting the AFL directly is quite complicated. A simple solution how to set up the AFLs for all zones in *(k,q)* semidual space is described by Alg.7.1.

**Construction of AFL algorithm:**

    **for** k:=1 **to** N **do** (* N is number of polyhedron facets *)
        **if** facet$_k$ interferes with the zone[4] (i,j) defined by corners (i,j) and (i+1,j+1)
        **then** add facet$_k$ into the AFL$_1$[i,j]; (* i,j are indexes in dual space *)
--

**Algorithm 7.1**

This algorithm is computed six times - for *xy*, *xz* and *yz* projections and every for both semidual representations - we compute AFL$_1$..AFL$_6$, see Table 6.1.

## 8. Theoretical considerations of O(1) algorithm in $E^3$

The proposed algorithm has been tested and compared with the CB algorithm as the CB algorithm is very stable and its behaviour does not depend on geometric properties of the given polyhedron and on clipped lines. Since the proposed algorithm is supposed to be superior over other modifications of CB algorithm it is necessary to make theoretical estimation of its efficiency. It is necessary to point out that algorithm efficiency can differ from computer to computer. For $5.10^7$ operations ( := , < , ± , * , / )  we get the following timing ( 33 , 50 , 16 , 20 ,114 ).

Let us assume that *N* is number of facets of given polyhedron. For algorithm efficiency considerations we will consider:

 - CB algorithm complexity, see [Ska93], can be described as
$$T_{CB} = (9,3,6,6,1)*N$$
    that is for considered timing
$$T_{CB} = 777*N$$

- proposed algorithm with $O(1)$ complexity is defined as
$$T_{O(1)} = (18,3,8,8,4) + T_{CB}(2)$$
    and using considered timing

---
[4] The „butterfly" zone and its semidual representations - figure 6.1

$$T_{O(1)} = 1488 + 1554 = 3042$$

The part „$T_{CB}(2)$" is uses of CB algorithm for determining points of intersection.
Let us introduce theoretical algorithm efficiency coefficients as:

$$\boxed{\nu_{1T}(1) = \frac{T_{CB}(1)}{T_{O(1)}} = 0.26} \qquad \boxed{\nu_{2T}(25) = \frac{T_{CB}(25)}{T_{O(1)}} = 6.39}$$

This value is **approximate**. We count with number of processing lines but some parts are computed only for intersecting or nonintersecting lines. Direction and position are important too. We reason theoretical number of facets in final set W about 5 (see alg. 3.1), but in theoretical consideration we reason the best possibility - only two facets in final set W.

## 9. Experimental results of O(1) algorithm in $E^3$

Five types of O(1) algorithm tests were made . Some interesting results and graphs are included in text.

*Test 1*

The first test is determining the speed of the algorithm with a various number of the polygon facets in dependence on subdivision of dual space. In table and in some graphs there adequate results of CB algorithm are presented . For even type of subdivision of dual space three graphs are presented. Preprocessing graph, processing graph and sum of preprocessing and processing consuming time graph. Very interesting is the third graph. This graph shows an optimal subdivision of the dual space. The shown optimum is relative. It is optimum for sum of preprocessing and processing time, but in some applications we want to minimize the time of processing. For this application type, the presented optimum is improper.

Tables and graphs are included in appendix. The tests are computed for $n_k=n_q$, $n_k=2$, $n_k=5$, $n_q=5$ and $n_q=10$.

Here are presented two graphs. Fig 9.1 show summary results for polyhedron consisting of 2112 facets. Fig 9.2 show 3D image of processing time consuming.

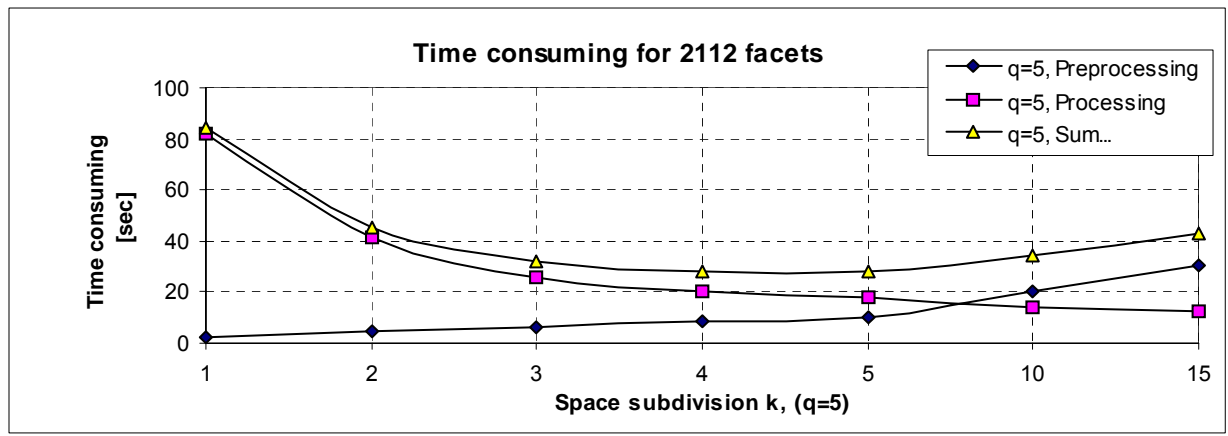

**Fig. 9.1.**

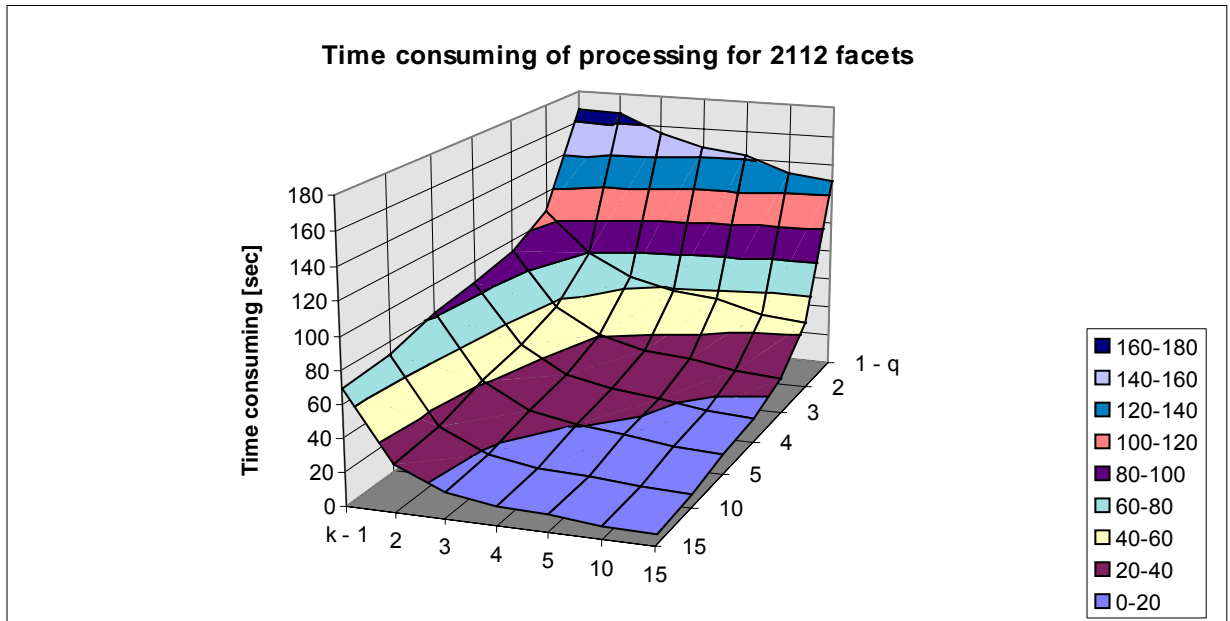

**Fig. 9.2.**

*Test 2*

Results of the second algorithm test shows dependence of speed on number of lines. It is important that comparation graphs (Fig. 9.3) show ratio $v_1 = \dfrac{T_{CB}}{T_{O(1)}}$ of algorithms. The ratio increases which means that algorithm O(1) is faster. Because this graph is fundamental, its copy is presented here (and in appendix too - part Test 2). Horizontal axis shows number of facets, vertical axis is logarithm of relation $v_1 = \dfrac{T_{CB}}{T_{O(1)}}$.

In graph we can see - if number of facets is greater than 24 (1000 lines are processed) - than O(1) algorithm have positive effect is faster than CB algorithm. For number of facets equal to 24 is $v_{1E} = 1$ (E = experimental). This number is fundamental for determining effectivity of O(1) algorithm. In graph we can see, that the effectivity is not linear. We can assume, that maximal of $v_1$ value is about 4:

$$\boxed{v_{1E} = \dfrac{T_{CB}}{T_{O(1)}} = 4}$$

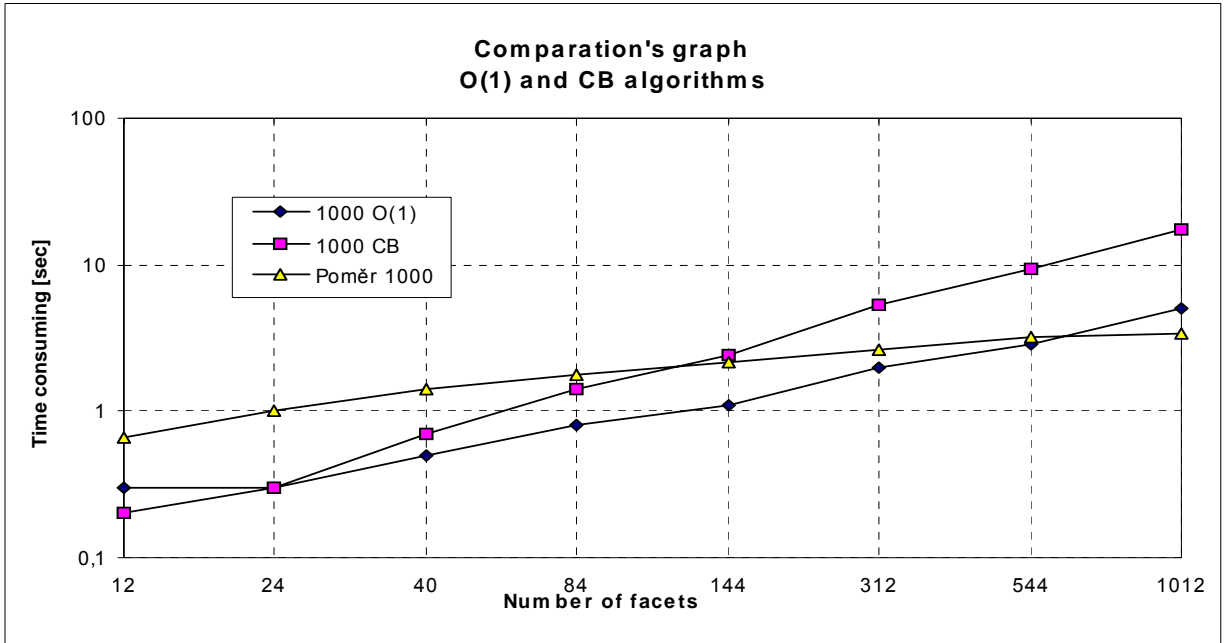

**Fig. 6.3.**

*Test 3*

The third test verified length of AFLs in dependence on subdivision of dual space. It was computed for various number of facets. The graphs (Fig. 9.4) present different influence of subdivision *k* and subdivision *q* on the length of AFL. The 3D graph shows it significantly.

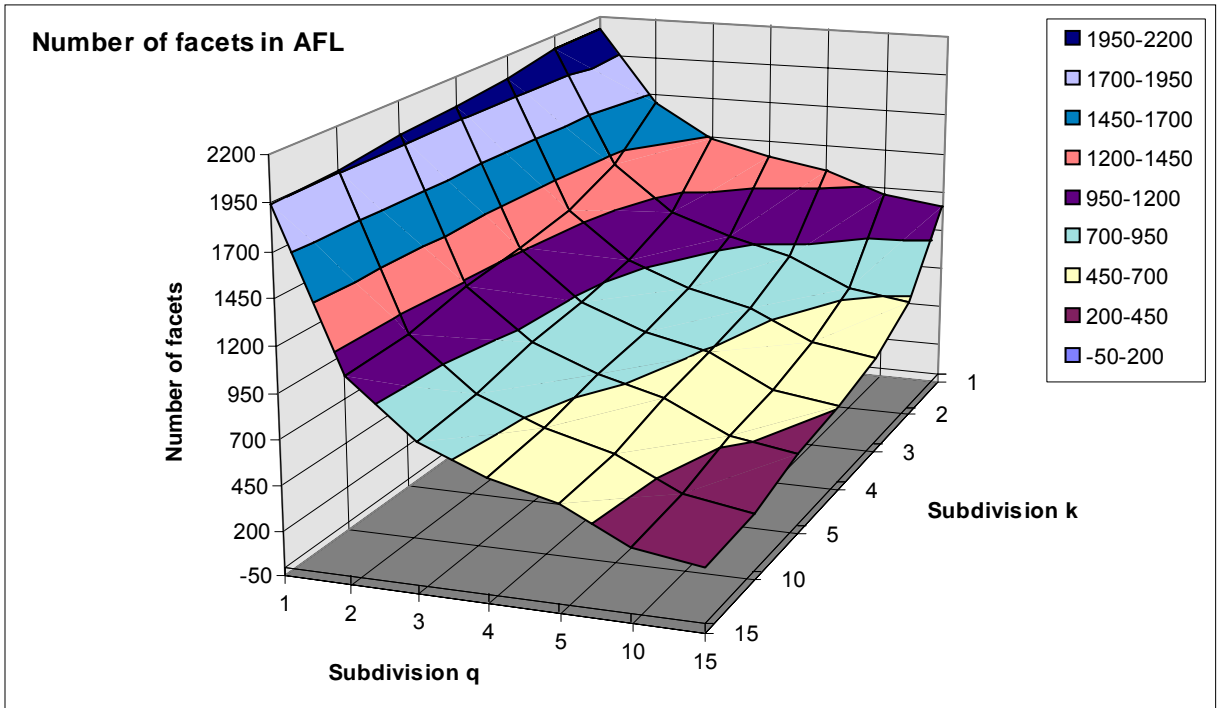

**Fig. 6.4.**

*Test 4*

The fourth test shows dependence of algorithm speed on number of intersecting and nonintersecting lines, see Fig. 9.5.

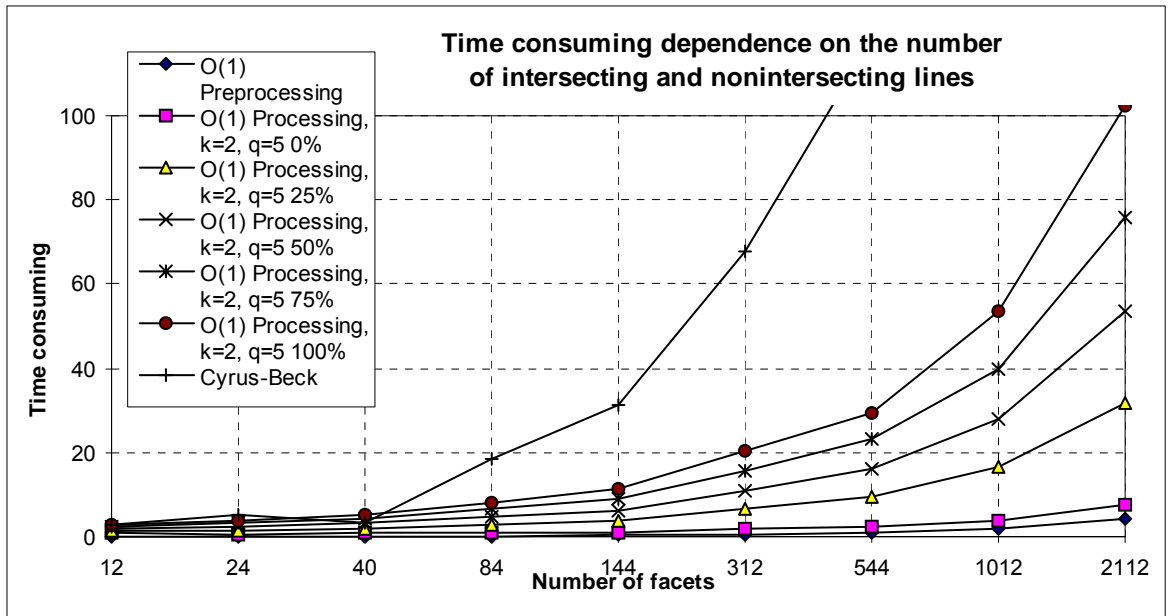

**Fig. 6.5.**

*Test 5*

The last test is only different expression of „Test 1". Fig. 9.6 show time consuming dependence on the number of facets. (4-4, 5-5 etc. are values of subdivision *k-q*)

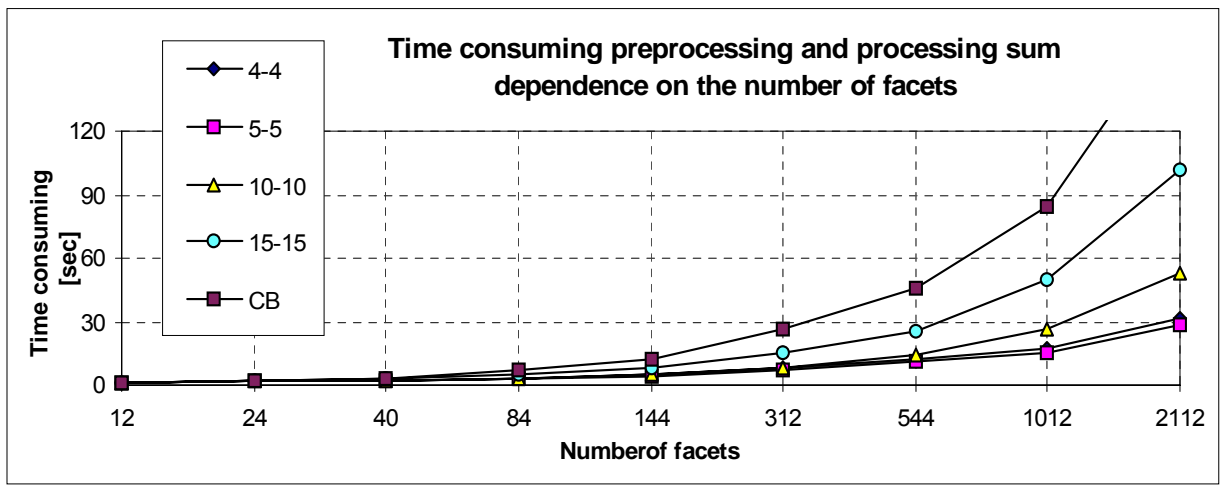

**Fig. 6.6.**

## 10. Conclusion

The new algorithm for line clipping by convex polygon in $E^2$ and/or polyhedron in $E^3$ was tested. Algorithms were compared to Cyrus-Beck line clipping algorithm. The algorithms are superior than the CB algorithm. The proposed algorithms are convenient for those applications where clipping area is stable and many lines are clipped. The algorithms claims the expected processing complexity $O(1)$.

The presented approach can be applied in many areas of computer graphics and there is a hope that it can be used to find new trends for trading space and speed.

All tests were implemented in C++ on PC 486 / 50 MHz.

All tests are stored in the appendix. Appendix is available on URL: http://herakles.zcu.cz or by e-mail.

## 11. Acknowledgements

Authors would like to express their thanks to Miss I. Kolingerová and to the students of Computer Graphics courses at the University of West Bohemia in Plzeò for their suggestions and critical comments that stimulated this project .